\documentclass{article}

\usepackage{graphicx}      
\usepackage[T1]{fontenc}   

\usepackage{natbib}
\usepackage{amsmath}
\usepackage{amssymb}
\usepackage{hyperref}
\usepackage{color}
\usepackage{float}
\usepackage{placeins}
\usepackage{stfloats}
\usepackage{subfig}

\usepackage{dsfont}
\newcommand{\ind}{\mathds{1}}

\usepackage{multirow}
\usepackage{enumitem}
\usepackage{url}
\usepackage{stackengine}
\usepackage{booktabs}

\usepackage[utf8]{inputenc}

\def\plaintitle{An Interpretable Produce Price Forecasting System for Small and Marginal Farmers in India using Collaborative Filtering and Adaptive Nearest Neighbors}

\begin{document}

\title{\plaintitle}

\author{Wei Ma$^{1}$, Kendall Nowocin$^{2}$, Niraj Marathe$^{2}$, George H.~Chen$^{1,2}$ \vspace{1em}\\
${}^1$Carnegie Mellon University \\
${}^2$CoolCrop}

\date{}




\maketitle

\begin{abstract}
Small and marginal farmers, who account for over 80\% of India's agricultural population, often sell their harvest at low, unfavorable prices before spoilage. These farmers often lack access to either cold storage or market forecasts. In particular, by having access to cold storage, farmers can store their produce for longer and thus have more flexibility as to when they should sell their harvest by. Meanwhile, by having access to market forecasts, farmers can more easily identify which markets to sell at and when. While affordable cold storage solutions have become more widely available, there has been less work on produce price forecasting. A key challenge is that in many regions of India, predominantly in rural and remote areas, we have either very limited or no produce pricing data available from public online sources.

In this paper, we present a produce price forecasting system that pulls data from the Indian Ministry of Agriculture and Farmers Welfare's website Agmarknet, trains a model of prices using over a thousand markets, and displays interpretable price forecasts in a web application viewable from a mobile phone. Due to the pricing data being extremely sparse, our method first imputes missing entries using collaborative filtering to obtain a dense dataset. Using this imputed dense dataset, we then train a decision-tree-based classifier to predict produce prices at different markets. In terms of interpretability, we display the most relevant historical pricing data that drive each forecasted price, where we take advantage of the fact that a wide family of decision-tree-based ensemble learning methods are adaptive nearest neighbor methods. We show how to construct heuristic price uncertainty intervals based on nearest neighbors. We validate forecast accuracy on data from Agmarknet and a small field survey of a few markets in Odisha.
\end{abstract}

\section{Introduction}

Agriculture accounts for 14\% of India's GDP and is the principal source of income for roughly 50\% of the Indian population \citep{jha2015report}. 
Small and marginal farmers, who have a field size of less than~2 hectares, make up over 80\% of India's agricultural population~\citep{agricensus}.\footnote{In India, a ``marginal'' farmer cultivates agrictultrual land up to 1 hectare, and a ``small'' farmer cultivates land from 1 to 2 hectares \citep{agricensus}. An overview of challenges and opportunities for small farmers in India is provided by \citet{mahendra2014small}.} Many of these farmers' incomes are heavily tied to profits from selling their produce. To increase profits, these farmers could try to sell their produce at higher prices. However, the issue is that if they set the produce prices too high, then they run the risk of having a lot of leftover unsold produce. This unsold produce can then spoil and become unfit for consumption. To avoid having unsold produce, farmers often sell their produce at low, unfavorable prices. To make matters worse, smallholder farmers often devote a large portion of their fields to traditional food crops while using only a small fraction of their fields for cash crops \citep{van2001off}. Two key problems are that small and marginal farmers often lack access to cold storage or to market information and forecasts.

Cold storage can keep produce fresh for longer before they spoil. 
Within the last decade, enormous progress has been made in increasing cold storage capacity and improving cold storage infrastructure in India \citep{nccd2015}.
For example, circa 2007, there were no cold storage facilities in Odisha or Uttar Pradesh \citep{umali2007food}. By 2014, Odisha had 111 cold storage facilities with a capacity of 326,639 metric tons, and Uttar Pradesh had 2,176 cold storage facilities with a capacity of 13.6 million metric tons \citep[Annexure~VI]{nccd2015}. The Indian government's Ministry of Agriculture and Farmers Welfare and Ministry of Food Processing Industries have both emphasized cold chain development, which includes cold storage. A lot of initial efforts focused on facilities tailored for a select few cross-seasonal crops such as potatoes. Small and marginal farmers often instead grow crops that have dramatically shorter storage times such as tomatoes. Moreover, these farmers might not be close to an existing cold storage facility and even if they are close to one, some of these facilities can have high costs. To provide affordable cold storage to such farmers, many companies are now producing relatively small cold storage units each with a capacity of a few metric tons that can be shared by a few dozen farmers. For example, there is the Ecofrost cold room by ecoZen\footnote{\url{http://www.ecozensolutions.com/innovation/micro-cold-storage}}, the Modular Cold Room by Unitech Engineering Solutions\footnote{\url{https://www.indiamart.com/unitechengineeringsolutions/}}, and
the ColdShed by SolerCool HighTech Solutions\footnote{\url{http://www.solercoolhightech.com/}}. This list is far from exhaustive. 

With small and marginal farmers increasingly having access to cold storage, a general problem becomes how to optimally use cold storage. Numerous related questions arise. How much should farmers store of which produce in cold storage? Where and when should they sell specific produce? How should the space be divvied up among the farmers sharing a cold storage unit? These sorts of questions could influence which cold storage solution farmers choose in the first place. Meanwhile, the actual process of selling produce can be lengthy and can involve a farmer driving out to a potentially far away market and staying at the market for a while to sell produce. Visiting multiple markets in one trip may make sense as well. How should farmers plan these routes and decide on how long they stay at each market? Do they need an additional cold storage unit on the vehicle? Farmers can better answer the above operations-related questions if they have reliable market information and forecasts. At least for soybeans, there is evidence that giving farmers direct access to market information enables them to sell soy at higher prices and improves their welfare~\citep{goyal2010information}. 



To help provide farmers with market information, many web and mobile apps have been recently developed as news and information content platforms. The Indian government's Ministry of Agriculture and Farmers Welfare provides a website with an accompanying mobile app called Agmarknet\footnote{\url{http://agmarknet.gov.in/}} that provides produce pricing and volume data spanning over a decade for over a thousand markets. Meanwhile, farmers can buy and sell agriculture-related products and services on Agribuzz, and FarmBee provides real-time produce prices, weather, and market news. In addition to market information, plant protection and agriculture advisories can be obtained using an app called Kisan Suvidha. 
Despite there being many apps that provide farmers with market information, we are unaware of any existing apps that focus on interpretable produce price forecasts for farmers.

\begin{figure}[!t]
	\centering
	\includegraphics[width=0.6\columnwidth]{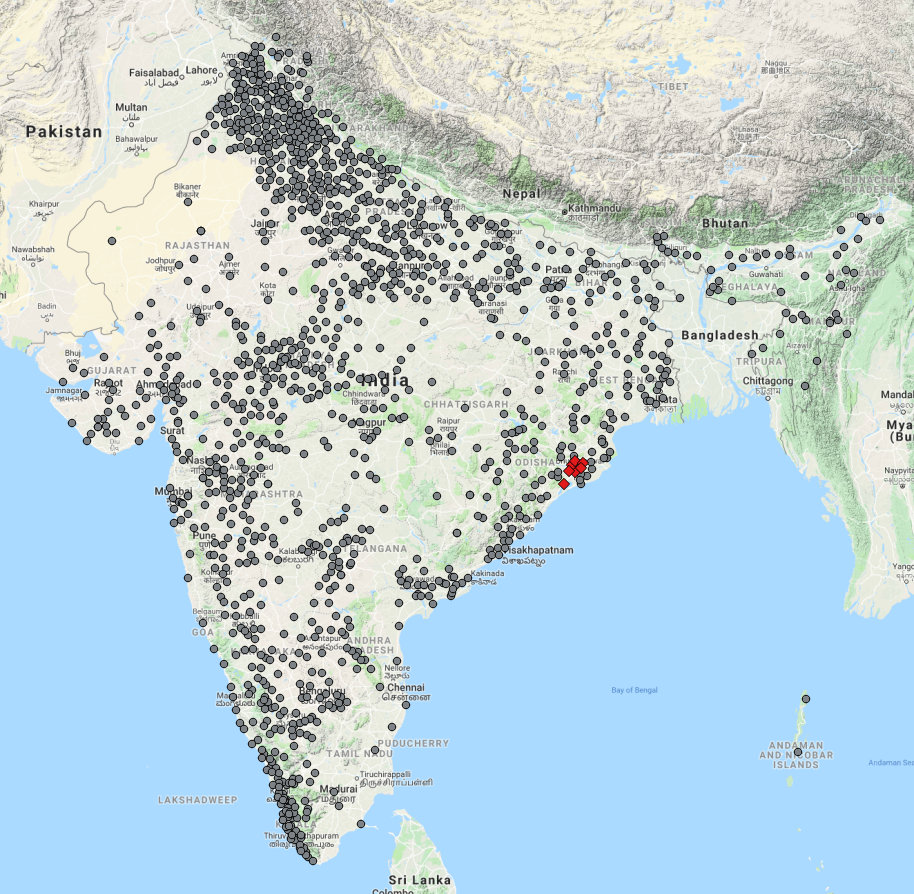}
	\vspace{-.5em}
	\caption{Location of markets from Agmarknet (grey dots) and the field survey (red diamonds).}\label{fig:map}
\end{figure}

As produce price forecasting somewhat resembles stock price forecasting, some studies view produce as special stocks with high volatility, low frequency, and low trading volume \citep{brandt1983price, kumar2004price, chen2017toward}. However, produce markets differ from stock markets in a few key aspects:
\begin{itemize}[leftmargin=*]
	\item Produce markets are located all over the country (see Figure~\ref{fig:map}), while stocks are traded at a centralized location through the internet. Produce markets in disparate geographical locations may have prices that are quite different for the same produce. Sales at markets can be by a loud outcry auction bidding \citep{goyal2010information}.
	\item Pricing and volume data for a produce at a market can be missing if the market is closed for the day, the produce is unavailable for the day, or the data simply does not get recorded despite the produce being sold at the market. Note that data is manually entered at each government-regulated market.
	\item Produce prices contain unexpected noise and outliers as a result of both price negotiations and a more error-prone manual data entry process.
	\item As previously discussed, ``executing a trade'' (i.e., a farmer selling produce) is more time and labor intensive than stock trading.
\end{itemize}
Overall, the distributed nature of produce markets and how data are collected and entered per market lead to produce pricing and volume data being noisy and extremely sparse. We provide more details on data scraped from Agmarknet as well as data manually collected through a field survey in Section \ref{sec:data}.

In this paper, we develop a system for forecasting produce price trends while providing evidence for forecasts. Our system uses collaborative filtering to impute missing data and adaptive nearest neighbor methods to obtain interpretable forecasts (Section~\ref{sec:pred}). For interpretability, we show which historical pricing information (which markets and which calendar dates) drive any particular forecasted price change, and we also construct heuristic uncertainty intervals for forecasted prices (Section~\ref{sec:nn}). We use the proposed forecasting system to predict price changes at 1352 markets in India for six produce (Section~\ref{sec:ana}). We also present a web app that updates and displays price forecasts on a daily basis (Section~\ref{sec:web}). This web app has already been deployed at a pilot site in Odisha. We are actively working with farmers on making the web application more useful. 

\paragraph{Related work} To the best of our knowledge, only a few studies work on produce markets. \citet{brandt1983price} explored time series models and expert models to predict produce prices for the next quarter, and \citet{chen2017toward}~predicted the price trend in the near future with a number of machine learning methods. \citet{chakraborty2016predicting}~predict food prices using both Agmarknet data and news events with the idea that some news events (e.g., worker strikes, festivals) can help predict price shocks. None of the above studies predict produce prices for multiple time periods and multiple markets, and their prediction results are not presented in a fashion that is easy to interpret and explore by farmers. 
The closest work to this paper is that of \citet{chen2017toward}, which benchmarks a few machine learning algorithms forecasting price change for the next day looking at 14 markets and one specific produce.

\section{Market Data Characteristics}
\label{sec:data}

\begin{figure}[!t]
	\hspace{-3.5em}
	\includegraphics[width=1.25\columnwidth]{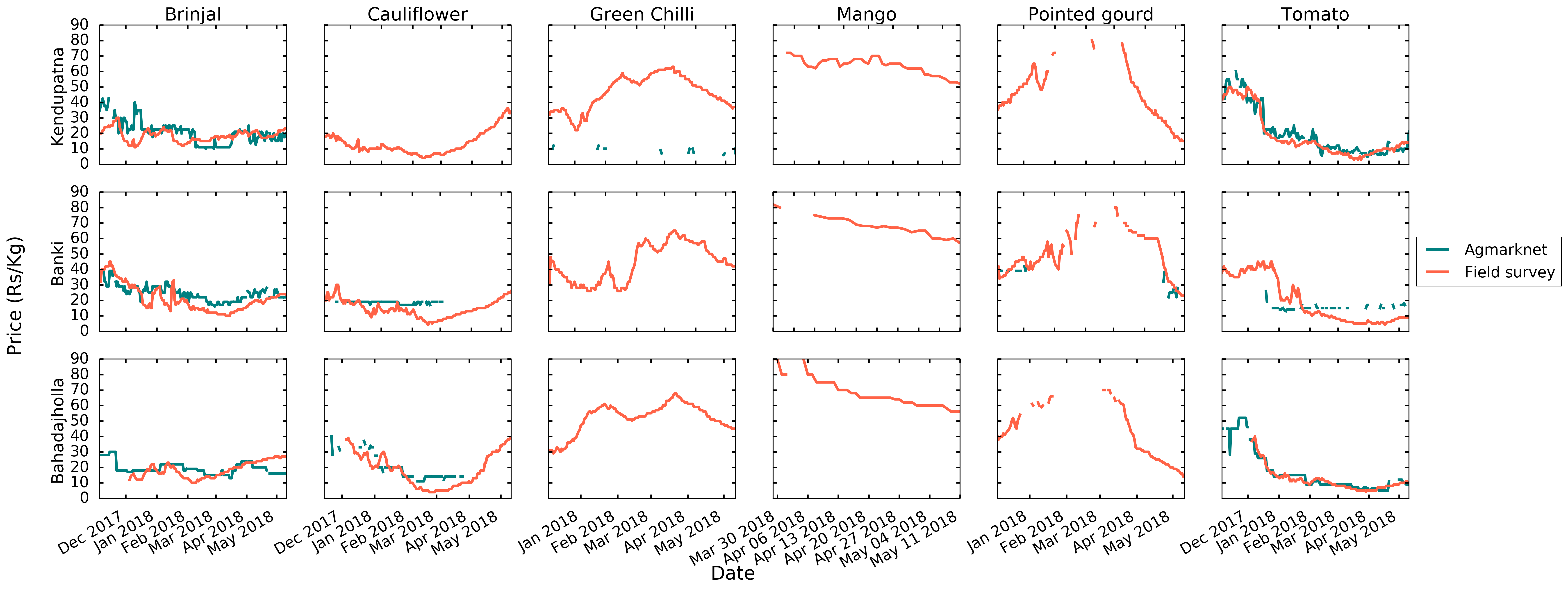}
	\vspace{-1.5em}
	\caption{Prices of six produce at three markets.}\label{fig:price}
\end{figure}

We collect produce pricing data from two sources: Agmarknet and a field survey. Agmarknet is run by the Indian government's Ministry of Agriculture and Farmers Welfare and contains price and volume data at 1352 markets across India for over twelve years. For the field survey, we hired an individual in India to collect price and volume data of six produce (brinjal, cauliflower, green chilli, mango, pointed gourd, tomato) at six markets in Odisha from November 26, 2017 to June 24, 2018. The data collection is conducted through phone calls and in-person communication with local retailers. Markets in Agmarknet are located all over India, whereas the field survey only covers markets near Cuttack in Odisha. 
The market locations from Agmarknet and the field survey are shown in Figure~\ref{fig:map}. Note that only three markets (Bahadajholla, Banki, Kendupatna) appear in both Agmarknet and the field survey. Our experiments later will focus on the six produce that appear in the field survey data and occasionally will also focus only on the three overlapping markets.

The two data sources contain a lot of missing entries, i.e., a specific produce at a specific market can have no recorded price or volume information across a large number of days. This missingness in data is caused by a market being closed, a produce not being sold on a particular day at a market, or the price or volume data for a produce at a market simply not being recorded even though the produce was indeed being sold and the market was open. We show the prices of the six produce over time at the three overlapping markets for the two data sources in Figure~\ref{fig:price}. For Agmarknet data in Figure~\ref{fig:price}, 
the range of dates for which pricing data are available is different among different produce, and the prices of some produce (e.g., pointed gourd) are missing over the entire time period. The field survey data has better coverage for the six produce recorded than Agmarknet data at the Sunakhela market; however, brinjal, cauliflower, and tomato prices are still missing before January 1st, 2018. Agmarknet rarely has green chilli and mango prices at the three markets while prices are still available in the field survey data.

Although prices from Agmarknet and the field survey usually follow similar trends, they still have some differences. For both data sources, data are collected via surveying farmers and prices are manually recorded. The two data sources survey different sets of farmers, so the prices collected between the two data sources are not expected to match. We remark that the prices we consider for both data sources are the modal price per produce for each day (when pricing data are actually available for that particular day). Meanwhile, we do not know what the rate of error is in verbal communication of prices or in data entry. 

Lastly, we mention two characteristics observed in Agmarknet data that inform how we set up our forecasting system in the next section. We remark that the pricing information shown in Figure~\ref{fig:price} (corresponding to November 26, 2017 to June 24, 2018) is actually the test data in our numerical experiments (Section~\ref{sec:ana}). As we do not have field survey data prior to November 26, 2017, we only use Agmarknet data up to November 25, 2017 for training forecasting models. Specifically in examining this training Agmarknet data for the six produce of interest, we find that the prices of these different produce tend to trend differently and spike at different times. As a result, we train separate forecasting models for different produce. Furthermore, as prices vary between markets, each forecasting model is specific to a particular market. However, we observe in the training Agmarknet data that prices of markets in close geographic proximity tend to have similar prices. Thus, when we train a forecasting model for a specific market, the training data we use will come from that market as well as nearby markets.


\section{Forecasting Method}
\label{sec:pred}

In this section, we present our approach to forecasting produce prices. For simplicity, we train a different model per produce, so throughout this section, we assume that there is a single produce of interest. As input, we take pricing information (of the single produce of interest) from all Indian markets up to present time. As output, we forecast the direction of price change (up, down, or stay the same) per market at the next ``time step''. What constitutes a ``time step'' is based on how time is quantized. We present results later where one time step corresponds to 1 day, 1 week, 2 weeks, or 4 weeks. Our forecasting approach can easily be modified to predict exact prices at the next time step (regression) rather than just the direction of price changes (classification). However, we emphasize that predicting exact prices is a much harder problem than predicting the direction of price changes. For clarity of exposition, we introduce our method in the classification context. We defer discussing the regression context to Section~\ref{sec:regression}.

\begin{figure}[!tp]
	\centering
	\includegraphics[width=0.77\columnwidth]{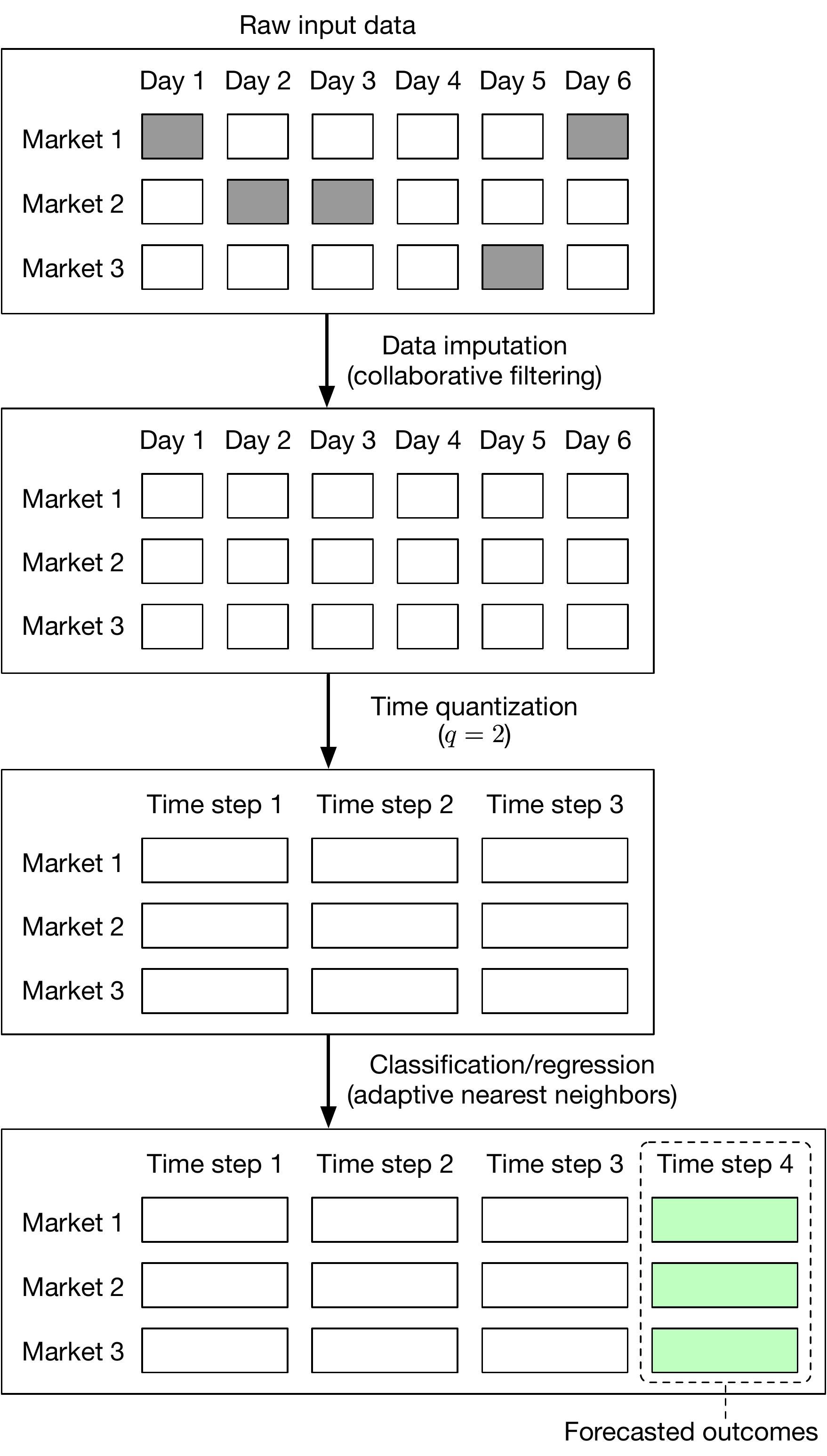}
	\caption{Overview of forecasting system (white boxes represent pricing data, shaded boxes represent missing data, and green boxes represent predicted outcomes).}\label{fig:frame}
\end{figure}

Our forecasting method consists of three steps. First, we impute missing data using a standard collaborative filtering method; at this point, we use the finest level of granularity for time in the raw input data. In the second step, we quantize time. Lastly, using the dense imputed data that have been quantized in time, we train a model using an adaptive nearest neighbor method such as random forests or boosting methods that use decision trees as base predictors. A pictorial overview of these three steps is shown in Figure~\ref{fig:frame}. We explain these three steps in detail in the next three subsections.

We remark that for the third step, we intentionally use adaptive nearest neighbor methods because they can readily provide ``evidence'' for forecasts \cite[Section 7.1]{chen2018explaining}. In our problem context, this means that for any predicted price change direction, an adaptive nearest neighbor method can provide historical dates and prices that are directly used in making the prediction; these historical dates and prices can be supplied to the farmer as forecast evidence and, as we discuss in Section~\ref{sec:nn}, can also be used to construct a heuristic ``uncertainty interval''. This uncertainty interval can be thought of as a range of plausible prices associated with a forecasted price change direction. Instead of adaptive nearest neighbor methods, other machine learning classifiers could be used instead although providing forecast evidence and some notion of a price uncertainty interval may be difficult depending on the method used.


\subsection{Data imputation}

\begin{figure}[!t]
	\subfloat[Raw\label{fig:bimpute}]{\includegraphics[width=0.9\columnwidth]{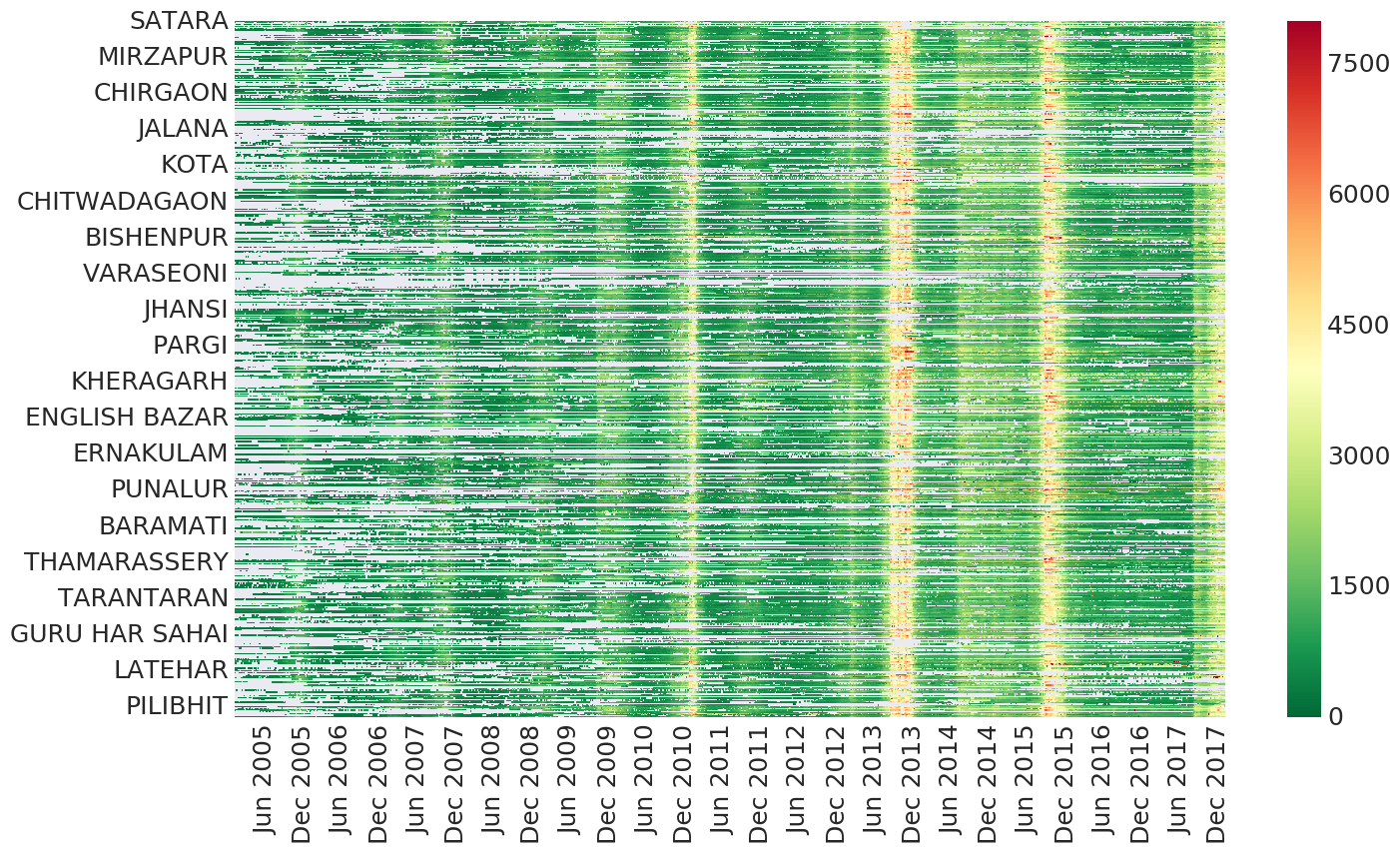}} \vspace{-.5em} \\
	\subfloat[Imputed\label{fig:impute}]{\includegraphics[width=0.9\columnwidth]{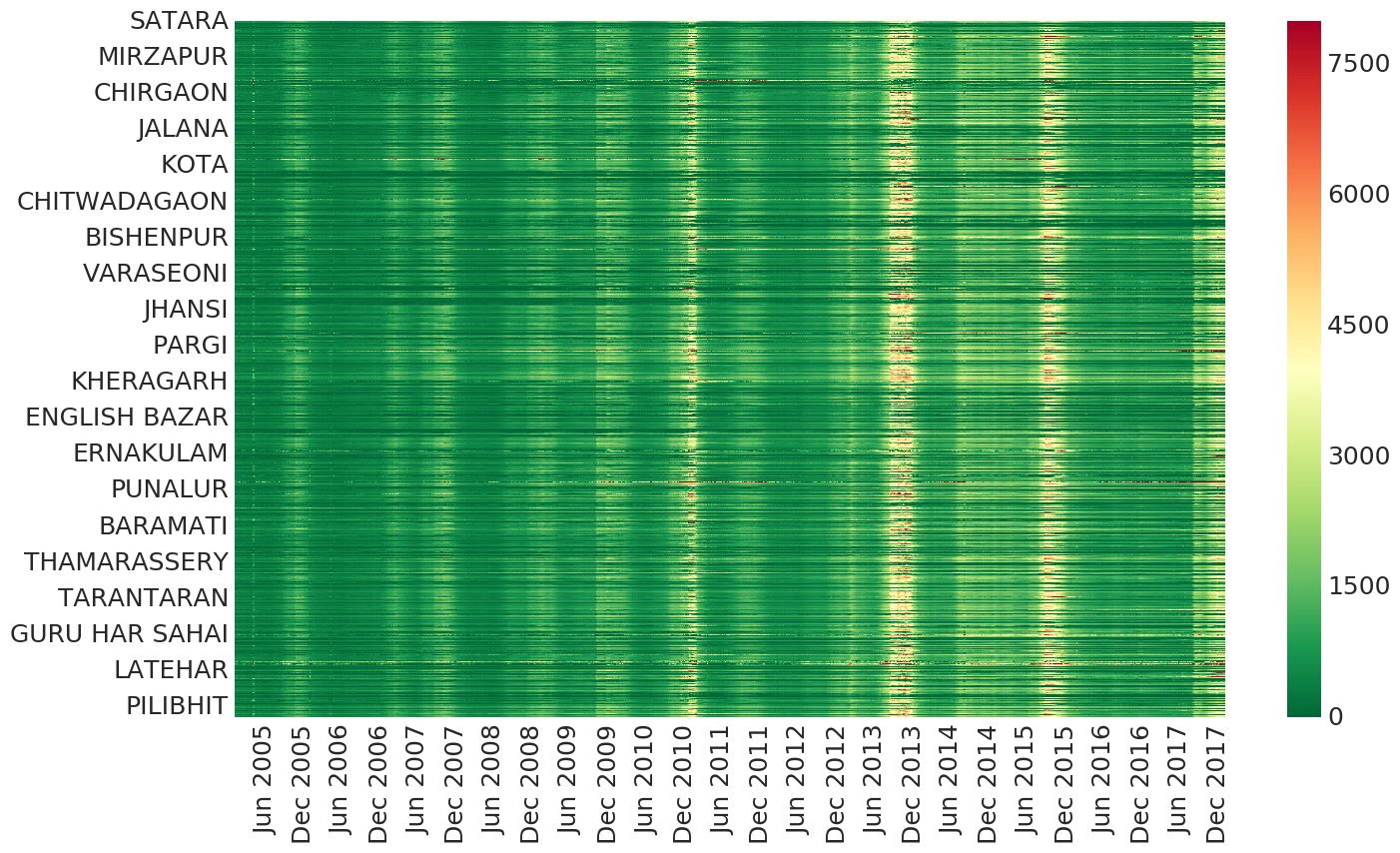}}
	\caption{Onion prices for all Agmarknet markets (Rs/100 kg). In each heat map, while all markets are shown, only the names of 20 markets are displayed on the left side.}\label{fig:before-after-impute}
\end{figure}

We denote $M$ as the number of markets and $T$ as the number of consecutive calendar days in the training data. Recall that we are considering a single produce. We define $P$ to be the $M$-by-$T$ price matrix for the produce of interest, where $P_{m,t}$ denotes the produce price at market $m\in\{1,2,\dots,M\}$ on day $t\in\{1,2,\dots,T\}$; if this price is missing in the training data, then $P_{m,t}=\text{``NaN''}$. Using onion as an example, the price matrix $P$ is visualized as a heat map in Figure~\ref{fig:before-after-impute}\subref{fig:bimpute}. We see that onion pricing data are dense after 2015 and sparse before 2007. Onion prices tend to increase over the years, likely due to inflation. Meanwhile, onion prices exhibit seasonality structure, reaching a local maximum around October every year.

Standard collaborative filtering methods can be used to fill in the missing entries of matrix $P$, and we use 
the SVD-based \textsc{SoftImpute} algorithm by \citet{hastie2015matrix} in this paper. 
We denote the dense, imputed price matrix as $\widetilde{P}$, which now has no ``NaN'' entries. An example of imputed onion prices is shown in Figure~\ref{fig:before-after-impute}\subref{fig:impute}.



\subsection{Time quantization}
\label{sec:pb}


After imputation, we now have a dense price matrix $\widetilde{P}\in\mathbb{R}^{M\times T}$. At this point, we quantize time as follows. Let positive integer $q$ be the number of days we want to consider as a single time step. Then for every non-overlapping consecutive block of $q$ days, we average the produce prices within each block of $q$ days to obtain a time-quantized version of dense price matrix $\widetilde{P}\in\mathbb{R}^{M\times T}$; if $q$ does not divide $T$, then we ignore the last block that is shorter than $q$ days. We denote the resulting time-quantized dense price matrix as $Q\in\mathbb{R}^{M\times \lfloor T/q \rfloor}$, where $\lfloor\cdot\rfloor$ is the floor function. In particular, for $m\in\{1,2,\dots,M\}$ and $s\in\{1,2,\dots,\lfloor T/q \rfloor\}$,
\begin{align*}
Q_{m,s}
= \frac{1}{q} \sum_{t=s q}^{(s+1)q-1} \widetilde{P}_{m,t}.
\end{align*}

\subsection{Classification}
\label{sec:classification}

Now that we have quantized time so that a single time step corresponds to an average of $q$ prices, we set up a classification problem for predicting each market's price change direction at the next time step. To do so, we specify the feature vectors and corresponding labels that we train a classifier with. We begin by explaining how we obtain the classification labels.

We define the relative price change $\Delta_{m,s}$ at time step $s$ at market~$m$ as
\begin{equation*}
\Delta_{m,s}
= \begin{cases}
0 & \text{if }s = 1,
\\
\frac{Q_{m,s} - Q_{m,s-1}}{Q_{m, s-1}}& \text{otherwise}.
\end{cases}
\end{equation*}
Then the price change direction $y_{m,s} \in \{-1, 0, 1\}$ at time step~$s$ at market $m$ is defined as
\begin{equation}
y_{m,s} = \begin{cases}
-1 & \texttt{if $\Delta_{m,s} < 0$}, \\
0& \texttt{if $\Delta_{m,s} = 0$}, \\
1& \texttt{if $\Delta_{m,s} > 0$}.
\end{cases} \label{eq:classification-labels}
\end{equation}
These price change directions are precisely the labels that we use for classification. We have not yet described what their corresponding feature vectors are.

We construct feature vectors as follows. The basic idea is that we want to predict $y_{m,s}$ using information at time steps prior to time step~$s$. We look at information in the last $\tau$ time steps, where $\tau>0$ is a user-specified parameter; using a larger value of $\tau$ means that we incorporate information from further back in time in making predictions. A simple approach would be to simply use the relative price changes at these $\tau$ most recent time steps, i.e., $\Delta_{m,s-\tau-1}$, $\dots$, $\Delta_{m,s-1}$. We find that incorporating volume information (i.e., how much of the produce is being sold) from the most recent $\tau$ time steps is also helpful. Let $V_{m,s}$ denote the volume of the produce of interest being sold at market~$m$ at time step~$s$ (as with pricing information, we can impute missing volume data via collaborative filtering and then discretize time). Then for label $y_{m,s}$, we define its corresponding feature vector of length $2\tau$ to be
$$x_{m,s}=(\Delta_{m,s-\tau-1},\dots,\Delta_{m,s-1},V_{m,s-\tau-1},\dots,V_{m,s-1}).$$
Lastly, since markets that are close by geographically tend to have similar price trends, we further pull information from nearby markets. Specifically, for any market~$m$, let $\mathcal{M}_{k}(m)$ denote the $k$ markets closest in geographic proximity (Euclidean distance) to market $m$ (including market $m$). Then we train a classifier using feature vector/label pairs $(x_{m',s}, y_{m',s})$ for $s\in\{\tau+1,\dots,\lfloor T/q \rfloor\}$ and for every market $m'\in\mathcal{M}_{k}(m)$. Parameter $k$ is chosen later via cross validation. As mentioned previously, we favor using classifiers that are adaptive nearest neighbor methods as they readily provide forecast ``evidence'' and also yield a heuristic method for producing price uncertainty intervals.

At test time, to forecast a price change direction, we assume that we have access to the last $\tau$ time steps of pricing and volume information. From these most recent $\tau$ time steps, we construct a test feature vector $x_{\text{test}}$ in the same manner in which we construct training feature vectors. We denote the classifier learned as $\widehat{y}_m$, which takes a feature vector as input and outputs a price direction in $\{-1,0,1\}$ (note that the classifier is specific to both the market $m$ as well as a produce of interest; as a reminder, throughout this section we focus on a single produce of interest). Thus, $\widehat{y}_m(x_{\text{test}})$ is the predicted price change direction for feature vector $x_{\text{test}}$.

\subsection{Forecast evidence and price uncertainty intervals via adaptive nearest neighbors}
\label{sec:nn}

Training an adaptive nearest neighbor model corresponds to learning a similarity function $\mathbb{K}$ such that for any two feature vectors $x$ and $x'$ (constructed as in Section~\ref{sec:classification}), $\mathbb{K}(x,x')\in[0,1]$ is an estimated similarity between $x$ and $x'$. Explicit formulas for this learned similarity function $\mathbb{K}$ for decision trees and their bagging and boosting ensemble variants (such as random forests, gradient tree boosting, and AdaBoost with tree base predictors) are given by Chen et al~\cite[Section 7.1]{chen2018explaining}. When an adaptive nearest neighbor method makes a prediction for a test feature vector $x_{\text{test}}$, the prediction is computed using standard weighted nearest neighbors, where weights are given by the learned similarity function~$\mathbb{K}$. To be precise, for our classification setup in Section~\ref{sec:classification} and for any given test feature vector $x_{\text{test}}$, 
the estimated probability that $x_{\text{test}}$ has price change direction $d\in\{-1,0,1\}$ is
\begin{equation}
\widehat{\eta}_m(d|x_{\text{test}})
= \frac{ \sum_{m'\in\mathcal{M}_k(m)} \sum_{s=\tau+1}^{\lfloor T/q \rfloor} \mathbb{K}(x_{m',s}, x_{\text{test}}) \ind\{y_{m',s} = d\} }
        { \sum_{m'\in\mathcal{M}_k(m)} \sum_{s=\tau+1}^{\lfloor T/q \rfloor} \mathbb{K}(x_{m',s}, x_{\text{test}}) },
\label{eq:estimated-posterior-probability}
\end{equation}
where $\ind\{\cdot\}$ is the indicator function that is~1 when its argument is true and 0 otherwise; in fact the denominator is equal to 1. Then the adaptive nearest neighbor method's predicted price change direction for test feature vector $x_{\text{test}}$ is given by
\begin{equation}
\widehat{y}_m(x_{\text{test}})
= \underset{d\in\{-1,0,1\}}{\text{argmax}} \widehat{\eta}_m(d|x_{\text{test}}).
\label{eq:bayes-plug-in-classifier}
\end{equation}
Importantly, in computing this predicted price change for $x_{\text{test}}$, only training feature vector/label pairs $(x_{m',s},y_{m',s})$ with nonzero weight $\mathbb{K}(x_{m',s}, x_{\text{test}})$ contribute to the prediction. Concretely, we define the following set of ``nearest neighbors'' of $x_{\text{test}}$:
\begin{align*}
\mathcal{N}(x_{\text{test}})
&= \big\{ (m',s) \;:\; m'\in\mathcal{M}_k(m)\text{ and }s \in \{1,2,\dots,\lfloor T/q \rfloor \} \\
&\qquad\quad\quad\quad\; \text{ s.t.~}\mathbb{K}(x_{m',s}, x_{\text{test}}) > 0 \big\}.
\end{align*}
For example, if $(5, 42)$ is in $\mathcal{N}(x_{\text{test}})$, then it means that market 5's $\tau$ time steps immediately before time step 42 have information that contributes to the prediction for $x_{\text{test}}$ and this information has relative importance given by the weight $\mathbb{K}(x_{5,42}, x_{\text{test}})$. Note that we can figure out exactly which $q \tau$ calendar dates these $\tau$ time steps immediately before time step 42 correspond to. Thus, for any test feature vector $x_{\text{test}}$, we can determine precisely the historical pricing information (which markets and which calendar dates) that contributes to predicting the price change for $x_{\text{test}}$; the relative importance weights give a sense of which price windows matter more than others. These historical price trends and importance weights form the forecast evidence specific to test feature vector~$x_{\text{test}}$.

\paragraph{Heuristic price uncertainty intervals}
Test feature vector $x_{\text{test}}$ corresponds to some time step $s_{\text{test}}$ that we aim to predict a price change direction for. We can also construct a heuristic uncertainty interval for what the exact produce price will be at time step $s_{\text{test}}$ using the training feature vectors with nonzero similarity to $x_{\text{test}}$ (the nearest neighbors of $x_{\text{test}}$). Such an interval is heuristic in that we are unaware of any statistical guarantees on these intervals.

The idea is simple: each nearest neighbor $(m',s)\in\mathcal{N}(x_{\text{test}})$ of $x_{\text{test}}$ corresponds to a time window (the $\tau$ time steps before time step $s$ for market $m'$) that has been deemed by the classifier to be ``close'' to the $\tau$ time steps corresponding to the test feature vector $x_{\text{test}}$. Hopefully the historical price $Q_{m',s}$ is also close to the test feature vector's exact price at time step $s_{\text{test}}$. In particular, we can think of $Q_{m',s}$ as a guess for the price at test time step $s_{\text{test}}$, where this guess has an associated weight $\mathbb{K}(x_{m',s},x_{\text{test}})$.

At this point, there are various ways in which we can construct a heuristic interval of plausible produce prices at time step $s_{\text{test}}$. For example, we could choose all predicted prices which have a weight that is at least some user-specified threshold $\omega>0$, and then take the uncertainty interval to go from the minimum to the maximum of these predicted prices, i.e., we construct the interval $[L(x_{\text{test}}), U(x_{\text{test}})]$, where
\begin{align*}
L(x_{\text{test}})
&= \min_{\substack{(m',s)\in\mathcal{N}(x_{\text{test}}) 
   \text{~s.t.~}\mathbb{K}(x_{m',s},x_{\text{test}}) \ge \omega}} Q_{m',s}, \\
U(x_{\text{test}})
&= \max_{\substack{(m',s)\in\mathcal{N}(x_{\text{test}}) 
   \text{~s.t.~}\mathbb{K}(x_{m',s},x_{\text{test}}) \ge \omega}} Q_{m',s}.
\end{align*}
Smaller values of $\omega$ correspond to wider uncertainty intervals. Alternatively, we can use this same idea except where we only consider all predicted prices corresponding to the largest $\ell$ weights for some user-specified integer~$\ell$. Higher choices of $\ell$ correspond to wider uncertainty intervals. In both cases, $\omega$ and $\ell$ could, for instance, be chosen so that in training data, for some desired fraction of the time $\alpha$, the price at the test time step $s_{\text{test}}$ lands in the uncertainty interval with frequency at least $\alpha$.

\subsection{Forecasting exact prices instead of price change directions}
\label{sec:regression}

We now describe how our forecasting approach can easily be modified to forecast exact prices (regression) instead of price change directions (classification). We replace the discrete classification labels $y_{m,s}\in\{-1,0,1\}$ given by equation \eqref{eq:classification-labels} with continuous exact prices as labels, namely $y_{m,s}^{\text{regress}} = Q_{m,s}\in\mathbb{R}$. Instead of training a classifier, we train an adaptive nearest neighbor regression method (decision trees and their bagging and boosting ensemble variants can readily be used for regression instead of classification and are still adaptive nearest neighbor methods). After doing this training, we have a learned similarity function $\mathbb{K}$ and instead of equations~\eqref{eq:estimated-posterior-probability} and \eqref{eq:bayes-plug-in-classifier}, the forecasted price for a test feature vector $x_{\text{test}}$ is given by a single equation:
\begin{equation}
\widehat{y}_m^{~\text{regress}}(x_{\text{test}})
= \frac{ \sum_{m'\in\mathcal{M}_k(m)} \sum_{s=\tau+1}^{\lfloor T/q \rfloor} \mathbb{K}(x_{m',s}, x_{\text{test}}) y_{m',s}^{\text{regress}} }
        { \sum_{m'\in\mathcal{M}_k(m)} \sum_{s=\tau+1}^{\lfloor T/q \rfloor} \mathbb{K}(x_{m',s}, x_{\text{test}})}.
\label{eq:adaptive-nearest-neighbor-regression}
\end{equation}
Price uncertainty intervals can be constructed in the same manner as before.



\section{Experimental Results}
\label{sec:ana}

In this section, we validate our forecasting framework using Agmarknet and field survey data. For both data sources, we use six months of pricing and volume information as test data, from November 26, 2017 to June 24, 2018. We train forecasting models only using Agmarknet data before November 26, 2017 (the training also involves hyperparameter tuning via cross validation). Note that we do not have any field survey data available before November 26, 2017. Thus, we use the field survey data strictly for testing.

Our forecasting framework can be used with any classifier although we can only provide forecast evidence for adaptive nearest neighbor classifiers. We show the prediction accuracy on test data using one baseline classifier (multinomial logistic regression \citep{cox1958regression}), which is not an adaptive nearest neighbor method, and three adaptive nearest neighbor classifiers (random forests \citep{breiman2001random}, gradient tree boosting \citep{friedman2001greedy}, and AdaBoost with tree base predictors \citep{freund1997decision}) in Section~\ref{sec:selection}. We find random forests to be the most accurate among the four classifiers tested. We then show the accuracy of heuristic price uncertainty intervals (based on random forests) in Section~\ref{sec:int}. Lastly, we forecast the exact price using the proposed heuristic regression method of Section~\ref{sec:regression} and compare its accuracy with four standard regression models. 
In the following sections, all the models are tested on six produce (brinjal, cauliflower, green chilli, mango, pointed gourd, tomato) and we set $\tau = 10$.



\subsection{Classifying price change direction}
\label{sec:selection}

We apply our forecasting framework using four different classifiers: multinomial logistic regression, random forests, gradient tree boosting, and AdaBoost with tree base predictors. In the tables to follow, we abbreviate these four classifiers as ``LR'', ``RF'', ``GB'', and ``AB'', respectively. We use two accuracy measures: (1)~the raw accuracy, which is calculated as the fraction of $y_{m,s}$ that is correctly classified, and (2) the ``balanced'' accuracy \citep{chen2017toward}, which is calculated as the average of the raw accuracies in classifying each of the three price change directions $y_{m,s}\in\{1,0,-1\}$ (i.e., we calculate the raw accuracy confined to each of the three price change directions and then we average these three fractions).

We treat the pricing and volume information before November 26, 2017 as training data. Hyperparameters of the four classifiers are tuned via cross-validation for time series. In particular, the cross validation score for a specific classifier and choice of hyperparameters is computed as follows. First, we specify a beginning date $T_1$ and an end date $T_2$ for which we compute validation scores (in our experiments, we set $T_1$ to be Jan 1, 2017 and $T_2$ to be November 25, 2017). For each date $t$ ranging from $T_1$ to $T_2$, we train the classifier using the information before $t$ and compute its raw and balanced accuracies on date $t$. The validation score is taken to be the sum of the raw and balanced accuracies. Then the cross-validation score for the current classifier and choice of hyperparameters is calculated as the average validation score across all dates $t$ between $T_1$ and $T_2$. We sweep over the following parameters in the cross-validation: maximum rank for \textsc{SoftImpute}, number of closest markets $k$ to gather training data from per market (defined at the end of Section~\ref{sec:classification}), inverse of regularization strength for logistic regression, number of trees for random forests, and learning rates for both gradient tree boosting and AdaBoost. 


After running the above cross-validation procedure, we then train each classifier using all training data and using the best hyperparameters found from cross-validation. We then use these four classifiers to predict price change directions in the test data for Agmarknet and the field survey data. For both data sources, the test data are from November 26, 2017 to June 24, 2018. For different choices of time step sizes $q=1,7,14,28$, we report raw and balanced accuracies on the Agmarknet test data in Table~\ref{tab:res1}, and on the field survey test data in Table~\ref{tab:res2}. For Agmarknet data, each accuracy is the average over six produce and over 1352 markets across India. For the field survey data, each accuracy is the average over six produce and over three markets (Bahadajholla, Banki, Kendupatna). Note that the field survey data cover only a range of six months and are quite sparse, resulting in not enough data to use a time step size of $q=28$ days. For both datasets, random forests tend to outperform the other three classifiers in terms of both raw and balanced accuracy. 
Logistic regression consistently has the worst raw and balanced accuracies across the four classifiers tested. While gradient tree boosting and AdaBoost with tree base predictors can have performance competitive with and occasionally better than random forests, they do not consistently perform as well as random forests. 

\begin{table}[t]
\centering
\begin{tabular}{l c c c c c c c c}
 \toprule
 ~& \multicolumn{4}{c}{\small{\textbf{Raw Accuracy}}} & \multicolumn{4}{c}{\small{\textbf{Balanced Accuracy}}}\\
 \cmidrule(r){2-9}
 {\small \textit{Methods}}
 & {\small\textit{1}}
 & {\small \textit{7}}
 & {\small \textit{14}}
 & {\small \textit{28}}
 & {\small \textit{1}}
 & {\small \textit{7}}
 & {\small \textit{14}}
 & {\small \textit{28}} \\
 \midrule
LR & 0.65 & 0.60 & 0.57 & 0.64 & 0.39 & 0.36 & 0.41 & 0.49 \\
RF & \textbf{0.71} & \textbf{0.63} & \textbf{0.63} & \textbf{0.68} & \textbf{0.57} & \textbf{0.50} & \textbf{0.53} & \textbf{0.57} \\
GB & 0.69 & \textbf{0.63} & 0.61 & \textbf{0.68} & 0.55 & 0.49 & 0.52 & 0.56 \\
AB & \textbf{0.71} & \textbf{0.63} & 0.61 & 0.65 & 0.51 & 0.46 & 0.51 & 0.54 \\
 \bottomrule
\end{tabular}
\vspace{.5em}
\caption{Agmarknet test data raw and balanced accuracy for time step sizes $q = 1, 7, 14, 28$. For each column, in bold are the best performer(s) among the four classifiers tested.}\label{tab:res1}
\end{table}

\begin{table}[t]
\centering
\begin{tabular}{l c c c c c c c c}
 \toprule
 ~& \multicolumn{4}{c}{\small{\textbf{Raw Accuracy}}} & \multicolumn{4}{c}{\small{\textbf{Balanced Accuracy}}}\\
 \cmidrule(r){2-9}
 {\small \textit{Methods}}
 & {\small\textit{1}}
 & {\small \textit{7}}
 & {\small \textit{14}}
 & {\small \textit{28}}
 & {\small \textit{1}}
 & {\small \textit{7}}
 & {\small \textit{14}}
 & {\small \textit{28}} \\
 \midrule
 LR & 0.56 & 0.35 & 0.30 & N/A & 0.52 & 0.35 & 0.30 & N/A \\
 RF & \textbf{0.76} & 0.44 & \textbf{0.67} & N/A & \textbf{0.60} & 0.42 & \textbf{0.67} & N/A \\
 GB & 0.74 & \textbf{0.46} & 0.60 & N/A & 0.55 & \textbf{0.48} & 0.60 & N/A \\
 AB & \textbf{0.76} & 0.41 & 0.50 & N/A & 0.59 & 0.40 & 0.50 & N/A \\
 \bottomrule
\end{tabular}
\vspace{.5em}
\caption{Field survey test data raw and balanced accuracy for time step sizes $q = 1, 7, 14$. For each column, in bold are the best performer(s) among the four classifiers tested.}\label{tab:res2}
\end{table}



Because random forests tend to perform better than the other three classifiers for both Agmarknet and survey data, in the subsequent sections, when we produce forecast evidence and price uncertainty intervals, we only present results for random forests. Moreover, in the web application that we developed, we also only provide forecasts to farmers based on random forests.


\subsection{Calibrating price uncertainty intervals}
\label{sec:int}

\begin{figure}[t]
	\centering
	\includegraphics[width=0.7\columnwidth]{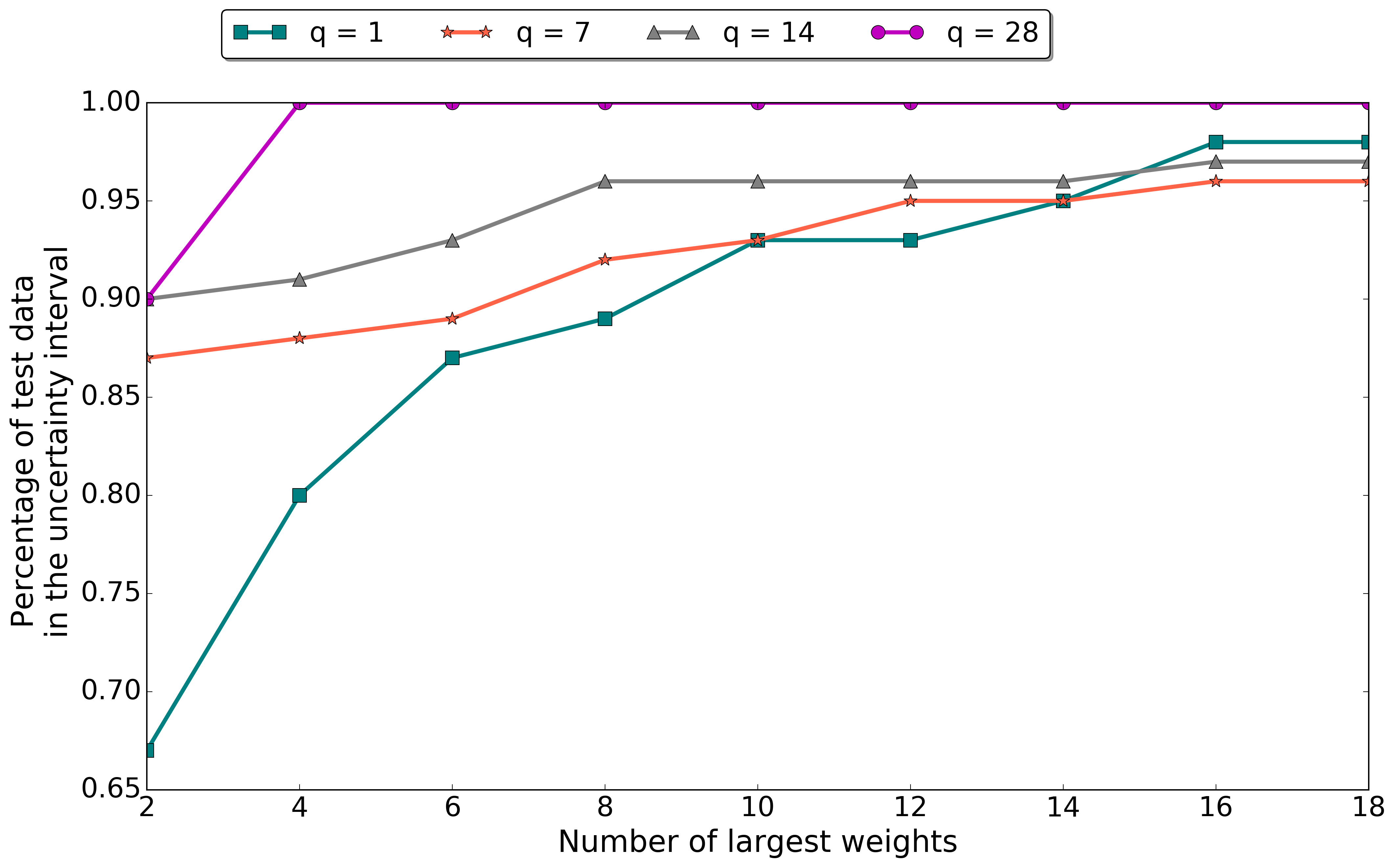}
	\vspace{-1em}
	\caption{Percentage of test data in the uncertainty interval.}\label{fig:nn}
\end{figure}

\begin{figure}[t]
	\centering
	\includegraphics[width=0.7\columnwidth]{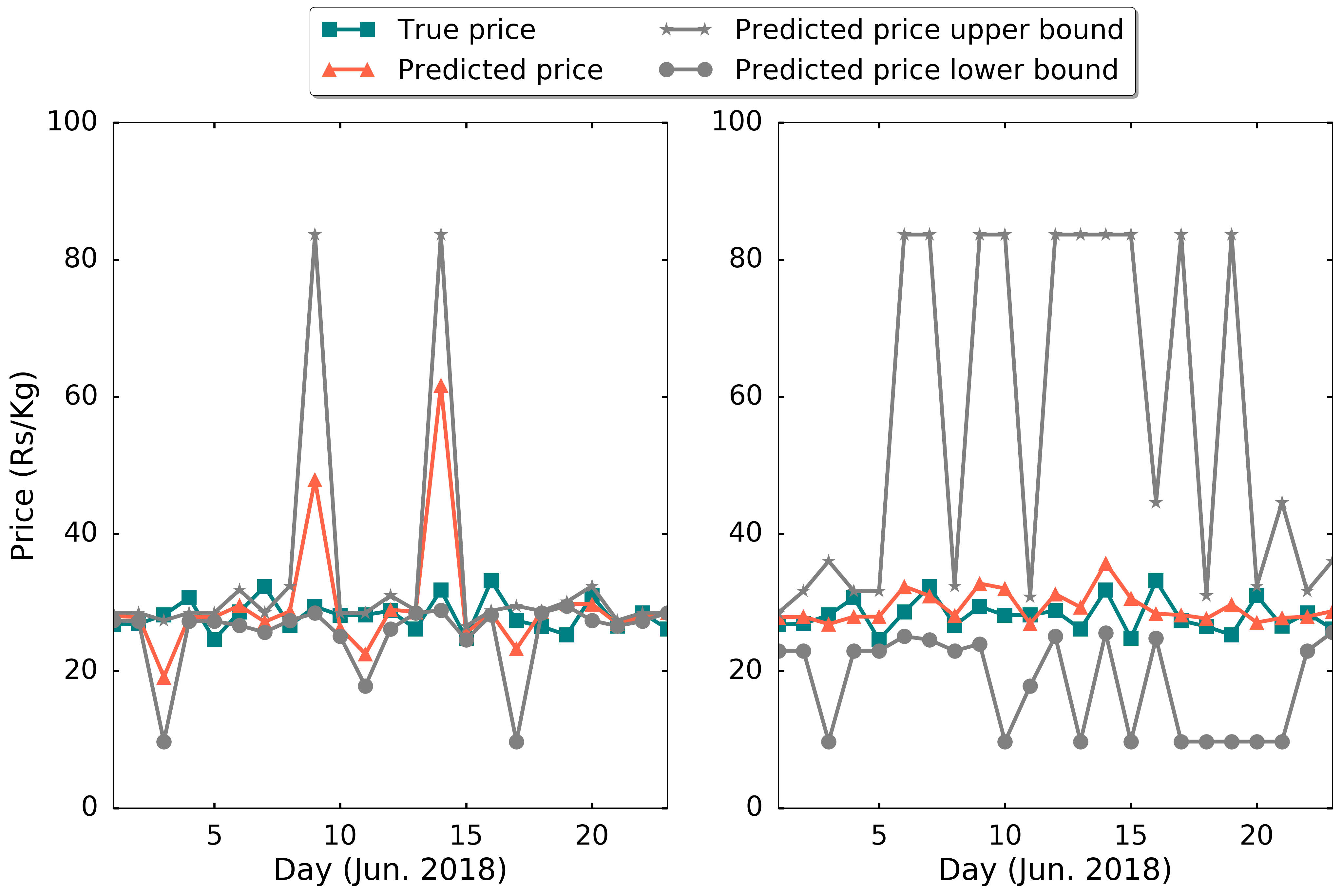}
	\vspace{-1em}
	\caption{Examples of the uncertainty intervals for mango price with step size $q=1$ (left: $\ell =2$, right: $\ell = 28$).}\label{fig:nn2}
\end{figure}

We now present experiments for constructing price uncertainty intervals and evaluating their accuracy, i.e., how often do our heuristic price uncertainty intervals actually contain the true price. We use random forests trained in Section~\ref{sec:selection} and extract the nearest neighbors $\mathcal{N}(x_{\text{test}})$ for the test Agmarknet data. The price uncertainty intervals are constructed based on the most similar $\ell$ adaptive nearest neighbors as presented in Section~\ref{sec:nn}. We vary $\ell$ and for each choice of $\ell$, we compute the percentage of price uncertainty intervals that contain the true prices that they are aiming to bound. For simplicity, in this section, these percentages are averaged over the same six produce that we have focused on throughout the paper and also over the three markets that are in both Agmarknet and the field survey data. The results are shown in Figure~\ref{fig:nn}. Specifically for mangoes in the market Banki, we show the price uncertainty interval over time (upper and lower bounds) vs the true price for two choices of $\ell$ in Figure~\ref{fig:nn2}.

Increasing $\ell$ means that we use more adaptive nearest neighbors to construct each price uncertainty interval. As a result, the intervals become wider and thus more easily contain the true price being predicted. Of course, a wider uncertainty interval is less useful to the farmer as it suggests that we are less certain in the prediction. For a large time step size of $q=28$ days, changes in price end up being small; consequently, prices becomes easier to predict and using a small choice for $\ell$ already results in price uncertainty intervals that easily contain the true price. Depending on how accurate we want the price uncertainty intervals, we could use a plot like Figure~\ref{fig:nn} to decide on a choice for~$\ell$.

\subsection{Forecasting exact produce prices}
\label{sec:regress}

To forecast exact produce prices rather than only price change directions, we can use the regression method presented in Section~\ref{sec:regression}, where the similarity function is computed based on the random forest classifier learned in Section~\ref{sec:selection}. In the table to follow, we denote this method as ``RFNN''. Note that this regression method is different from standard random forest regression. RFNN and standard random forest regression are both adaptive nearest neighbor methods and can be written in the form of equation~\eqref{eq:adaptive-nearest-neighbor-regression}. The key difference is that when growing the decision trees, RFNN uses a classification objective to decide on splits whereas standard random forest regression uses a regression objective. We compare RFNN with four standard regression methods: linear regression (abbreviated as ``LiR''), random forest regression (``RFR''), gradient tree boosting regression (``GBR''), and AdaBoost with decision trees that do regression as base predictors (``ABR'').
The four regression methods are tuned and trained using the same cross-validation procedure described in Section~\ref{sec:selection}, and the only difference is that we change the prediction output from price change directions to exact prices. In comparing these four methods with RFNN, we use root-mean-squared-error (RMSE) to measure the distance between true and predicted prices. The results are presented in Table~\ref{tab:res3}. Note that each value in Table~\ref{tab:res3} is the average over the six produce used throughout the paper and the three markets in both Agmarknet and the field survey data.

\begin{table}[t]
\centering
\begin{tabular}{l c c c c}
 \toprule
 ~& \multicolumn{4}{c}{\small{\textbf{RMSE} (Rs/kg)}} \\
 \cmidrule(r){2-5}
 {\small \textit{Methods}}
 & {\small\textit{1}}
 & {\small \textit{7}}
 & {\small \textit{14}}
 & {\small \textit{28}} \\
 \midrule
 RFNN & 7.81 & 14.81 & 6.57 & 2.12 \\
 LiR  & 12.07 & 14.91 & 6.58 & \textbf{2.02}\\
 RFR  & \textbf{7.50} & 14.82 & 6.53 & 2.18 \\
 GBR  & 7.80 & 14.81 & 6.52 & 2.13 \\
 ABR  & 10.03 & \textbf{14.27} & \textbf{6.47} & 4.78 \\
 \bottomrule
\end{tabular}
\vspace{0.5em}
\caption{Agmarknet test data RMSE for time step sizes $q = 1, 7, 14, 28$.}\label{tab:res3}
\end{table}

RFNN, random forest regression, and gradient tree boosting regression achieve similar accuracies for all time step sizes. Compared to these three methods, linear regression performs poorly for time step size $q=1$ but has better performance for $q=28$, when prediction is a much easier problem. AdaBoost has low error for $q=7$ and $q=14$ but does quite a bit worse than all the other methods for $q=1$ and $q=28$. This table suggests that different prediction methods could be used depending on which time step size $q$ we care about. However, if one wants to use the same method regardless of time step size, then RFNN, random forest regression, and gradient tree boosting regression appear the most promising; all three of these are adaptive nearest neighbor methods for which we can provide forecast evidence.


\section{Web application}
\label{sec:web}

\begin{figure}[!tp]
	\centering
	\includegraphics[width=0.6\columnwidth]{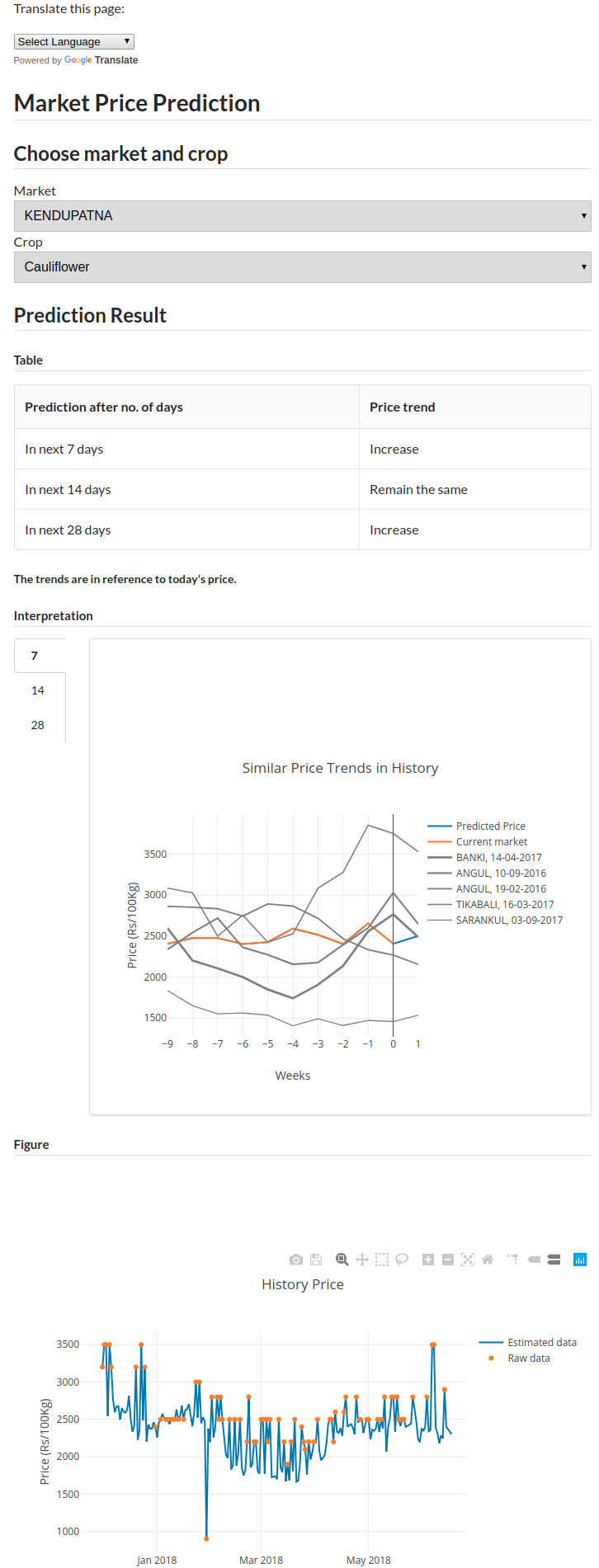}
	\caption{Screenshot of web application.}\label{fig:web}
\end{figure}

We now present a web application that acquires prices and updates price forecasts at $1352$ markets in India. The prediction results are updated daily and displayed on a web page. A screenshot of the web page is shown in Figure~\ref{fig:web}. The web application runs on an Amazon Web Services (AWS) server and includes a front end to display prediction results as well as a back end that routinely pulls data from Agmarknet, re-trains prediction models, produces forecasts, and sends various reports to server administrators. The technical specifications are listed in Table~\ref{tab:tech}.
 
\begin{table}[t]
	\centering
	\begin{tabular}{c c c c c}
		\toprule
		 \multicolumn{2}{c}{\small{\textbf{Front End}}} & \multicolumn{3}{c}{\small{\textbf{Back End}}}\\
		\cmidrule(r){1-5}
		{\small\textit{JavaScript}}
		& {\small \textit{CSS}}
		& {\small \textit{Server}}
		& {\small \textit{Database}}
		& {\small \textit{Database ORM}}\\
		\midrule
		jQuery & Semantic UI & Tornado & MySQL &  peewee \\
		\bottomrule
	\end{tabular}
	\vspace{0.5em}
	\caption{Technical specifications of the web application.}\label{tab:tech}
\end{table}

The back end runs the following sequential tasks daily at 8:00 PM Indian Standard Time:
\begin{enumerate}[label=\arabic*),leftmargin=*]
	\item {\em Acquire data:} download the current day's pricing data from Agmarknet.
	\item {\em Clean:} remove outliers.
	\item {\em Predict:} predict price change directions for the next day using our system described in Section~\ref{sec:pred}.
	\item {\em Archive data:} store the raw and imputed pricing data along with prediction results in the database.
	\item {\em Check and update:} check if the above processes are complete. If yes, update displayed results on the front end; otherwise, report issues to server managers.
	\item {\em Report:} send a report to server managers about the server status, quantity of acquired data, and prediction status.
\end{enumerate}
The whole process takes 3 to 4 hours to run over all produce and markets, and the updated forecasting results will be available at the beginning of the next day. In addition to the main forecasting process, a separate process is running periodically that re-trains and updates the prediction model per produce and per market with newly archived data.

For the front end, when users select the market and crop they want to query, the forecasting results for different time periods are displayed in a table. We also display the top most similar historical price change patterns (these are the nearest neighbors as described in Section \ref{sec:nn}). The web page is responsive in that it adapts to different screen sizes including most mobile phone screens. Since the vast majority of farmers involved in our pilot project do not read English but can read Hindi, we made sure to both incorporate manual translations for some price change directions and also incorporate a Google Translate add-on in the web page. We checked with our local partners to make sure that the translated pages are readable and understandable.

\paragraph{Pilot deployment and feedback from farmers}
We initiated a pilot project in Odisha with about 100 farmers at a single village. The pilot project can be potentially generalized to any area covered by Agmarknet, while the model localization and customization have to rely on the market data collected in each specific area. 
The pilot project is still ongoing.

The first round of feedback indicates that farmers want accurate exact price forecasts rather than price movement directions, and they want more produce listed on the web page (carrot, papaya, and pumpkin). 
Forecasting exact prices is more challenging than forecasting price movement directions. For forecasting exact prices, we prefer to provide some measure of uncertainty, which we can get with our price uncertainty intervals. As we discussed in Section~\ref{sec:int}, providing more accurate price uncertainty intervals means making the intervals wider, which also makes them less useful to farmers. We discuss some ideas for improving forecast accuracy in the next section. Meanwhile, we are changing the system to let the farmer customize what produce are shown. 
Most farmers have no issue accessing the webpage and most of them access it through a mobile phone. 
We are actively working with farmers to gather more user experience feedback.


\section{Discussion}
\label{sec:diss}

We have presented a fairly general forecasting framework. For the class of adaptive nearest neighbor methods, which includes a wide variety of existing classification and regression approaches, we further explain how to provide forecast evidence and construct heuristic price uncertainty intervals. However, we suspect that there are many ways to improve forecast accuracy and to also make the web app more useful to farmers. We discuss a number of future directions.


\subsection{Improving forecasts}

\paragraph{Additional features/datasets}
Our framework can easily incorporate more information than price and volume. We have obtained more datasets in India such as the Consumer Price Index (CPI), Producer Price Index (PPI), weather, and agricultural commodity prices from different data sources. This additional data could potentially increase the forecasting accuracy. \citet{chakraborty2016predicting} have also already demonstrated that using news data improves forecast accuracy of produce prices. 
A generalized data retrieval framework would need to be implemented to scrape the numerous additional datasets live and combine them into vector-valued time series data. Aside from this, model training and prediction remain the same.

\paragraph{Advances in prediction models} For data imputation, we also tested a novel imputation method called \textit{blind regression} \citep{song2016blind}. From preliminary experiments that we have run, replacing \textsc{SoftImpute} with blind regression results in improved prediction performance. However, it takes over one hour to complete the imputation for each produce. Due to computational constraints, we do not use blind regression in the web app. We are working on speeding up blind regression. Beyond the imputation step, advances in learning nearest neighbor representations could improve the overall prediction performance as well, for instance using \textit{differentiable boundary sets}~\citep{adikari2018efficient}.

\paragraph{Handling markets not on Agmarknet} For the field survey data, four markets in it are not covered by Agmarknet. The reason we collected data on them was that they are important to the farmers we are working with. However, we are unaware of any automated way to collect data from these four markets not on Agmarknet. More generally, many markets are not on Agmarknet and currently we do not know what is the best way to produce forecasts for these markets, unless of course there is some alternative online source that readily provides pricing and volume information for these markets.

A straightforward approach to handling these markets would be to task an individual or multiple people to manually collect pricing and volume information at these markets not easily accessible online. Then by looking at some initial collection of pricing and volume data, we figure out which markets on Agmarknet are most similar to these markets and use these similar Agmarknet markets to help forecast prices. What is unclear is how much data we need to manually collect before we can stop the manual collection at these markets. For some of these markets, we could potentially have to determine an automated way to keep collecting data.

\subsection{Improving usability of the web app}

\paragraph{Forecast interpretability} Presenting the forecasts in as user-friendly of a manner as possible remains a challenge. Right now, the way our forecast evidence is presented is still quite complex (the middle plot in Figure~\ref{fig:web}) and takes some time and explanation to parse. We are still figuring out how to improve the user interface and experience for farmers.

\paragraph{Web app accessibility}  The Internet is not always available for some rural areas in India, and cellular data is also costly. Most of the commercial cold storage systems include digital boards. We are now working with a provider of cold storage systems to incorporate the forecasting system on their digital boards. 

\subsection{Improving operations using forecasts}

\paragraph{Prediction reliability} For stock price prediction, investors can make money by the law of large numbers if predictions are correct ``on average''. However, farmers sell produce at a low frequency, and thus the price prediction needs to be more reliable. In addition to improving prediction accuracy, a matching and compensation mechanism can be developed to group farmers together and increase their ability to manage risks as a whole.

\paragraph{Decision making with predicted prices} Given the predicted price movements (or predicted prices/price intervals) and the cost of cold storage, optimal decisions can be made to maximize farmers' profits. To correctly formulate the optimization problem, we need to understand the constraints such as the maximum storage time and space requirements per produce. Most of the constraints come from farmers' experiences and preferences, so their input is extremely important in setting up an appropriate optimization problem to solve. This task can be challenging especially if farmers want to change their constraints and preferences over time.

\paragraph{Farmers' needs} We are also actively working with local farmers to understand their actual needs from the market data. For example, they may want to know the arrival volume of each produce such that they can decide which produce to sell. Some information such as when markets open and when during the day most sales happen might also be useful for farmers.


\subsubsection*{Acknowledgments}
This work was supported in part by a Carnegie Mellon University Berkman Faculty Development Fund. 


\bibliographystyle{ACM-Reference-Format}
\bibliography{predict}

\end{document}